\documentclass{elsarticle}

\usepackage{latexsym}
\usepackage{float}
\usepackage[latin2]{inputenc}
\usepackage{graphicx}
\usepackage{ae}
\usepackage{aecompl}
\usepackage{setspace}
\usepackage{amsmath}
\usepackage{amssymb}
\usepackage{amsthm}
\def\Pr{\hbox{\rm Pr}}

\def\AND{\hbox{\rm AND}}

\def\OR{\hbox{\rm OR}}

\usepackage[hidelinks]{hyperref}

\makeatletter
\def\ps@pprintTitle{%
	\let\@oddhead\@empty
	\let\@evenhead\@empty
	\def\@oddfoot{\centerline{\thepage}}%
	\let\@evenfoot\@oddfoot}
\makeatother

\begin{document}

\title{SCARF: A Biomedical Association Rule Finding Webserver}
	
\author[p]{Balázs Szalkai\corref{cor1}}
\ead{szalkai@pitgroup.org}
\author[p,u]{Vince Grolmusz\corref{cor1}}
\ead{grolmusz@pitgroup.org}
\cortext[cor1]{Corresponding authors}
\address[p]{PIT Bioinformatics Group, Eötvös University, H-1117 Budapest, Hungary}
\address[u]{Uratim Ltd., H-1118 Budapest, Hungary}

\date{}

	\begin{abstract}
		 The analysis of enormous datasets with missing data entries is a standard task in biological and medical data processing. Large-scale, multi-institution clinical studies are the typical examples of such datasets. These sets make possible the search for multi-parametric relations since from the plenty of the data one is likely to find a satisfying number of subjects with the required parameter ensembles. Specifically, finding combinatorial biomarkers for some given condition also needs a very large dataset to analyze. For this goal, statistical regression analysis is not the preferred tool of choice, since (i) the {\em a priori} knowledge of the parameter-sets to analyze is missing, and (ii) typically relatively few subjects have the interesting parameter-value ensembles for the analysis. For fast and automatic multi-parametric relation discovery association-rule finding tools are used for more than two decades in the data-mining community. Here we present the SCARF webserver for {\em generalized} association rule mining. Association rules are of the form: $a\ \AND\ b\ \AND\ ...\AND\ x \rightarrow y$, meaning that the presence of properties $a\ \AND\ b\ \AND\ ...\AND\ x$ implies property $y$; our algorithm finds generalized association rules, since it also finds logical disjunctions (i.e., ORs) at the left-hand side, allowing the discovery of more complex rules in a more compressed form in the database. This feature also helps reducing the typically very large result-tables of such studies, since allowing ORs in the left-hand side of a single rule could include dozens of classical rules.  The capabilities of the SCARF algorithm were demonstrated in mining the Alzheimer's database of the Coalition Against Major Diseases (CAMD) in our recent publication (Archives of Gerontology and Geriatrics Vol. 73, pp. 300-307, 2017). Here we describe the webserver implementation of the algorithm.
		
		\noindent{\bf Availability and implementation:}  The stand-alone SCARF (Simple Combinatorial Association Rule Finder) program is written in C++, and is downloadable from \url{https://pitgroup.org/apps/scarf/downloads/scarf.zip}. The webserver can be found at the address \url{https://pitgroup.org/scarf/}.
		
	\end{abstract}

	\maketitle
	
\section{Introduction and motivation} 

An enormous amount of data is generated every day in biological experiments and clinical investigations. These data may yield deep and very useful relations between parameters of interest if analyzed properly. Data mining techniques \cite{han-kamber,PDM}, which were first used in analyzing commercial transactions, are increasingly applied for biomedical data sources today \cite{Ivan2007,Ivan2009,Ivan2010}. Association rule mining is one of the areas that anticipated a massive development, beginning with its introduction in \cite{Assoc, Agrawal1994}. Association rules are automatically found patterns in large databases, where, say, each human patient has a number of attributes or parameter values, and the association rules describe implication-like relations between these attributes, like this one:
 \centerline{(high cholesterol level) AND (high blood pressure) $\to$ (heart disease).}
These rules have a left-hand side (abbreviated by LHS), left from the $\to$ symbol, and a right-hand side (RHS), right from the $\to$ symbol. There are several quality measures of these rules, we mention here only the three most important ones: 
 \begin{itemize}
 	\item Support: The number of data items (e.g., patients), where both the LHS and RHS are true. The LHS support is the number of the data items where LHS is true.
 	
 	\item Confidence: The value of the Support, divided by the LHS support. In our example it describes the fraction of patients with high cholesterol AND high blood pressure, also having heart disease.
 	
 	\item Lift: Describes the relative level of dependence between the LHS and the RHS, compared to the hypothesis that the LHS and the RHS are occurring independently; with probabilities: $\Pr(LHS \hbox{\rm ~AND } RHS)/(\Pr(LHS)\cdot\Pr(RHS))$. The lift is 1 if the LHS does not affect the RHS. The lift is greater than 1 if the LHS increase the probability of the occurrence of the RHS.
 \end{itemize}

In {\em association rule mining} the association rules with pre-defined minimum support, confidence and lift values need to be found \cite{Assoc, Agrawal1994}.

Here we present the SCARF algorithm and the related webserver that computes {\em generalized} association rules, where the LHS can also contain disjunctions (i.e., ORs), not only ANDs, as in the classical association rules. SCARF also computes some other statistical parameters of the rules. SCARF was successfully applied in mining the large CAMD Alzheimer's database \cite{Romero2009}, described in our work \cite{Szalkai2017b}.

The computed generalized association rules, with conjunctions and disjunctions in its LHS, have two remarkable properties: (i) any Boolean function can be represented as the ANDs of ORs of the variables and the negations of the variables, therefore, these generalized association rules are universal in describing Boolean functions, and (ii) short generalized association rules are capable of describing many non-generalized association rules in one formula, since, e.g., the LHS  $(a\ \OR\ b)\ \AND\ (c\ \OR\ d)\ \AND\ (e\ \OR\ f)$ is equivalent to the OR of eight ternary conjunctions; consequently, this generalized LHS compresses the LHS of eight non-generalized rules.

\section{Materials and methods}

The command-line SCARF program can be downloaded from \url{https://pitgroup.org/apps/scarf/downloads/scarf.zip}. This program takes a data table, a rule pattern and several numerical parameters (minimum confidence, etc.) as input, and produces association rules which have the given, pre-defined pattern.

The data table must be a comma or semicolon separated CSV file. The first line is the header, containing the column (or attribute or parameter) names. The subsequent lines are each a record in the data table, containing single ASCII characters in the cells. Empty cells are considered as N/A. Cells more than one character long are truncated to the very first character. Different characters represent different values.

The rule pattern is a logical expression, with blanks instead of variables. For example, $\square\ \AND\ (\square\ \OR\ \square) \to \square$ is a valid pattern. The allowed tokens in a valid rule pattern are blanks ($\square$), parentheses, operators (\texttt{AND}/\texttt{OR}) and the implies sign ($\to$). For the sake of simplicity, the right-hand side of a rule must always consist of a single blank.

SCARF examines all the sensible possibilities for filling the blanks in the rule pattern with elementary equalities. An elementary equality states that a database column (attribute) equals to one of some given values. This way we get the rule candidates. For example, $\mathrm{age}=AB\ \AND\ (\mathrm{bread=y}\ \OR\ \mathrm{onions=n}) \to \mathrm{butter=y}$ is a possible candidate for the pattern mentioned above. It states that if someone is in the age group A or B, and they buy bread or do not buy onions, then they will buy butter.

Numerical values like support, confidence, lift and leverage \cite{han-kamber} are assigned to all the candidates, then compared against the corresponding parameters which were supplied to the program, and rules which fail the test are discarded.

A significant time would be required if SCARF were to do an exhaustive search. Instead, branches of the rule tree are pruned in advance. If the first few blanks are filled a certain way, then in many cases we can deduce that we cannot get good enough rules by carrying on, no matter how we assign attributes to the remaining blanks. In this case, a backtrack happens earlier than it would if we tried to fill in the remaining blanks as well. This speeds up operation considerably.

Another improvement is the pooling of bit operations. A rule is a logical combination of some elementary equalities which are either true or false. So when we evaluate a rule, we have to perform logical operations on bits. We can speed up this part if we process multiple rows simultaneously. If the word-length of a computer is 64 bits, then we can process 64 rows in parallel, yielding a 64-fold improvement in running time.

The majority of processors have multiple cores today. If a program can utilize all the cores efficiently, it can achieve an $n$-fold speedup in the ideal case, where $n$ is the number of CPU cores. A commonly used approach is organizing a job pool and launch $n$ worker threads, which will be distributed to separate cores by the operation system. The worker threads will then pick jobs from the queue, process them and write back the result to another shared pool. If there are a lot of jobs and they are similarly computationally intensive, then this approach results in almost ideal resource distribution. That is, the threads will finish about in the same time and do not have to wait for each other in the end.

Since the right-hand side of the rule consists of a single blank, and we want to examine all the possibilities, each possible assignment of this blank can be regarded as a job for the threads. This means that SCARF creates as many jobs for the threads as the number of possible assignments for the blank on the right-hand side.

\section{Implementation and usage}

The SCARF webserver allows uploading data tables and setting parameters on a web form. The job is then run on our high-performance 16-core server, the typical running time is several minutes. After completion, the user is notified in an email and can view the results again on a web interface, both in TXT (unformatted) and XML (formatted). The first screen is for uploading the data table. There is a small example data table which can be used if the user just wants to test the service. It is possible to upload a custom data table which cannot exceed 2MB.

The next screen allows specifying the rule pattern, setting the parameters and constraining which columns (attributes) should appear on the left and right-hand side, respectively. It is advised to set stricter parameters to reduce the number of rules examined. The user can then schedule the job after checking the input once more.

The job is allowed 30 minutes of running time on our server. In our experiments, this was more than enough for a data table with 170 columns and 6100 rows. A user may submit five jobs daily, allowing two and a half hours of server-time per day. If a user would like to overcome these limitations, it is suggested downloading and running the off-line version.

\section*{Funding}
VG was supported by the VEKOP-2.3.2-16 program of National Research, Development and Innovation Office of Hungary.

\bigskip 
\noindent Conflict of Interest: The authors declare no conflicts of interest.

%\bibliography{v:/vince/CIKKEK/medl}
%\bibliographystyle{natbib}

\end{document}